\documentclass[11pt]{amsart}
\usepackage{graphicx}
\usepackage{subfigure}
\usepackage{amssymb,amsthm}
\usepackage{mathptm}
\usepackage{wrapfig}

\renewcommand{\paragraph}[1]{\smallskip\noindent{\bf #1}}

\newtheorem{thm}{Theorem}
\newtheorem{cor}{Corollary}

\newcommand{\W}{\mathcal{W}}
\renewcommand{\P}{\mathcal{P}}

\begin{document}

\title{Convex-Arc Drawings of Pseudolines}

\thanks{Parts of this work originated at Dagstuhl seminar 13151, \emph {Drawing Graphs and Maps with Curves}.
D.E. was supported in part by the National Science Foundation under grants 0830403 and 1217322, and by the Office of Naval Research under MURI grant N00014-08-1-1015. B.S. was supported by the Netherlands Organisation for Scientific Research (NWO) under project no.~639.023.208. M.v.G. and B.S. would like to thank Herman Haverkort for helpful discussions on the problem.}

\author[D. Eppstein]{David Eppstein}
\address{Computer Science Dept., University of California, Irvine, USA}
\email{eppstein@uci.edu}

\author[M. van Garderen]{Mereke van Garderen}
\address{Dept. of Computer and Information Science, University of Konstanz, Germany}
\email{mereke.van.garderen@uni-konstanz.de}

\author[B. Speckmann]{Bettina Speckmann}
\address{Dept. of Mathematics and Computer Science, TU Eindhoven, the Netherlands}
\email{b.speckmann@tue.nl}

\author[T. Ueckerdt]{Torsten Ueckerdt}
\address{Dept. of Mathematics, Karlsruhe Institute of Technology, Germany}
\email{ torsten.ueckerdt@kit.edu}

\keywords{Pseudolines, pseudoline arrangements, chord representations, wiring diagrams}

\begin{abstract}
A weak pseudoline arrangement is a topological generalization of a line arrangement, consisting of curves topologically equivalent to lines that cross each other at most once. We consider arrangements that are \emph{outerplanar}---each crossing is incident to an unbounded face---and \emph{simple}---each crossing point is the crossing of only two curves. We show that these arrangements can be represented by chords of a circle, by convex polygonal chains with only two bends, or by hyperbolic lines. Simple but non-outerplanar arrangements (non-weak) can be represented by convex polygonal chains or convex smooth curves of linear complexity.
\end{abstract}

\maketitle

\section{Introduction}

\subsection{Preliminaries}
Intuitively, \emph{pseudolines} are curves which behave like straight lines. This may be formalized in several ways;  we define a pseudoline to be what one gets from a line by stretching the plane without tearing it. In other words, it is the image of a line under a homeomorphism of the plane~\cite{shor1991stretchability}. Pseudolines extend to infinity in both directions, and cannot cross themselves, but their shapes have no additional restrictions. We consider here the problem of visualizing pseudoline arrangements using well-shaped curves.

A \emph{weak arrangement} of pseudolines is the partition of the plane induced by a set of pseudolines in which any two members of the set intersect at most once, and cross if they intersect~\cite{de2004stretching,goodman1993allowable}. A weak arrangement of pseudolines is an \emph{arrangement} of pseudolines if any two pseudolines in the set intersect exactly once. The intersection of one or more pseudolines in an arrangement is a \emph{vertex}. An \emph{ordinary vertex} is a vertex at which only two pseudolines intersect. An arrangement is \emph{simple} if all vertices are ordinary. We only consider simple arrangements. A vertex is a \emph{corner} if it is the most extreme (in one of the two directions) on both of its pseudolines. We primarily study weak arrangements that are \emph{outerplanar}. In an outerplanar arrangement (weak or non-weak), every crossing is part of an unbounded face of the arrangement.  

In Section \ref{sec:chords} we show that weak outerplanar pseudoline arrangements can be represented by a set of chords of a circle. To \emph{represent} in this case means to find a set of chords that is topologically equivalent to the arrangement. Furthermore, we show that we can also represent weak outerplanar pseudoline arrangements by convex polygonal chains (polylines) with at most two bends per chain, and by lines in the hyperbolic plane. In Section \ref{sec:arbitrary} we show that arbitrary (non-weak) pseudoline arrangements can also be drawn with convex polygonal chains, but may require a linear number of bends per chain. When we represent non-weak arrangement by smooth piecewise-circular curves, this may require $\Omega(n)$ arcs per curve.

\subsection{Related work}
One reason for interest in pseudolines is that they are in some ways more well-behaved than lines: determining whether a set of curves forms a pseudoline arrangement is trivial (just check whether each two curves cross at most once), while testing whether they are combinatorially equivalent to a line arrangement is NP-hard \cite{shor1991stretchability,Sch-GD-09}.
Pseudoline arrangements can be used to model sorting networks~\cite{AngHolRom-AM-07,Knu-AH-92}, tilings of convex polygons by rhombi~\cite{Sil-DM-93}, and graphs that have distance-preserving embeddings into hypercubes~\cite{Ber-EJC-08,Epp-EJC-06}. Pseudoline arrangements are also very closely related to oriented matroids, see e.g.~\cite{goodman2010}, and to combinatorial problems on planar points sets, see e.g.~\cite{ungar1982}.

Several results related to the visualization of pseudoline arrangements are known.
In \emph{wiring diagrams}, pseudolines are drawn on parallel horizontal lines, with crossings on short line segments that connect pairs of horizontal lines~\cite{Goo-DM-80}. Using a method based on compaction of wiring diagrams, the graphs of pseudoline arrangements may be given straight line drawings in small grids~\cite{Epp-GD-14}.
The planar dual graph of a weak pseudoline arrangement may be characterized as having drawings in which each bounded face is a centrally symmetric polygon~\cite{eppstein2005algorithms}. The pseudoline arrangements in which each pseudoline is a translated quadrant can be used to visualize \emph{learning spaces}, representing the possible states of knowledge of a human learner~\cite{eppstein2008upright}. Researchers in graph drawing have also studied force-directed methods for schematizing systems of curves representing metro maps by replacing each curve by a spline; these curves are not necessarily pseudolines, but they typically have few crossings~\cite{fink2013drawing}.

\section{Chord representations of outerplanar arrangements}
\label{sec:chords}

In this section, we show that there exists a \emph{chord representation} for every weak outerplanar pseudoline arrangement: a set of chords of a circle which are topologically equivalent to the arrangement. More precisely, if these chords are extended to lines, then there should exist an $\epsilon > 0$ such that intersecting the arrangement of the lines by the $(1+\epsilon)$-expanded disk produces an arrangement of line segments in the disk that is homeomorphic to the arrangement of pseudolines in the plane. If all corners are perturbed into the interior of the circle, then intersecting the chords with the disk itself (without any expansion) should again produce an arrangement homeomorphic to the pseudoline arrangement.

\begin{thm}
\label{thm:chords}
Every weak outerplanar pseudoline arrangement can be represented by a set of chords of a circle.
\end{thm}

\noindent\textit{Proof.}\quad Let $A$ be a weak outerplanar pseudoline arrangement. We prove, by induction on the number of crossings, the existence of a representation by chords of a circle in which each corner lies on the circle. Any or all corners may be perturbed into the interior of the circle without changing the other crossings. This is always possible because we are only dealing with simple arrangements, in which no more than two pseudolines can intersect at the same vertex.

\begin{figure}[tb]
\centering
\includegraphics{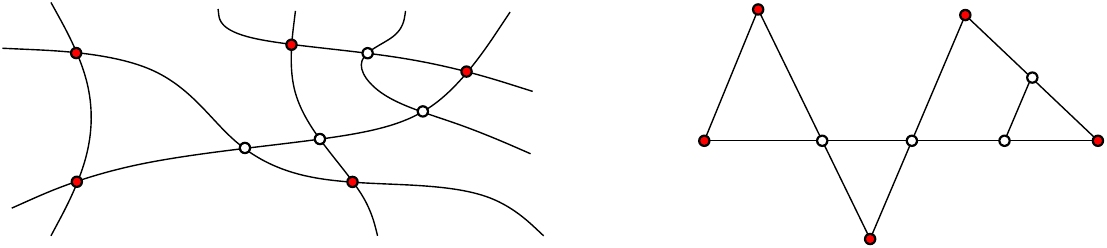}
\caption{Weak outerplanar pseudoline arrangement $A$ and its corresponding outerplane graph $G$, corner vertices are marked in red.}
\label{fi:arrangement}
\end{figure}

Let $G$ be the outerplane graph whose vertices are the vertices of $A$ and whose edges are segments of pseudolines between consecutive vertices in $A$. Figure \ref{fi:arrangement} shows a pseudoline arrangement $A$ and its corresponding graph $G$. Corners are shown in red, and non-corner vertices are shown in white. Note that $G$ has maximum degree four, because we only consider simple arrangements, and that no two triangles of $G$ share an edge, because that would indicate two pseudolines intersecting twice. One graph $G$ can correspond to multiple different pseudoline arrangements.

In the following, we describe how to represent $A$ by chords of a circle based on the properties of $G$. We distinguish six cases:
\begin{enumerate}

\item If $G$ is an isolated vertex, edge, or triangle, $A$ can be represented by chords of a circle as shown in Figure 2. Note that in these cases all vertices are corners, which we place on the circle. Each pseudoline is represented by exactly one chord.
\begin{figure}[htb]
\centering
\includegraphics{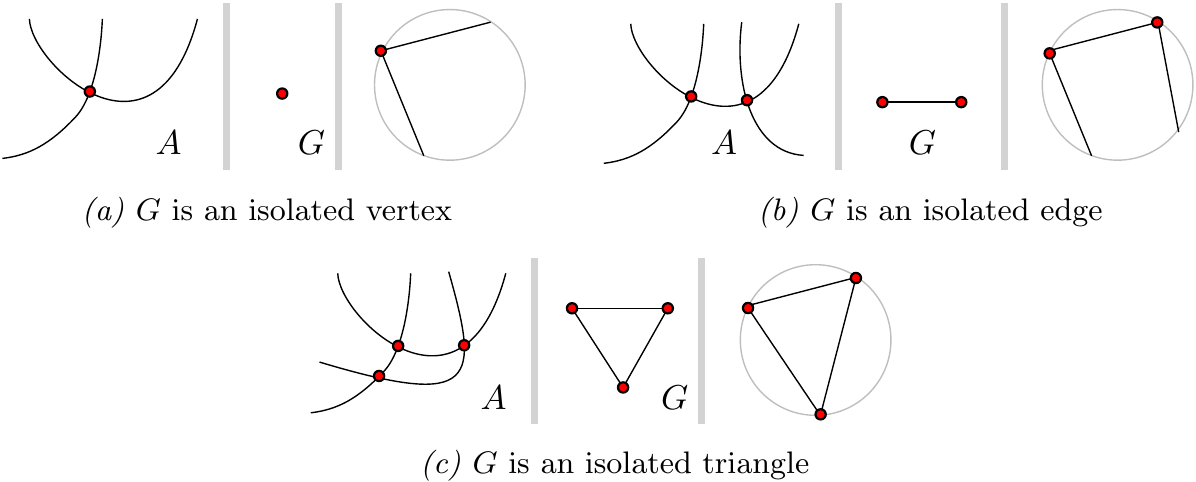}
\vspace{0.2em}
\caption{Chord representations for pseudoline arrangements where $G$ is (a) an isolated vertex, (b) an isolated edge, or (c) an isolated triangle. All vertices are corners, placed on the circle. Each pseudoline corresponds to exactly one chord.}
\label{fi:case1}
\end{figure}

\item If $G$ has a vertex $v$ of degree one, let $G'=G-v$. If $v$ in $G$ has degree one, one of the pseudolines crossing at the corresponding vertex $v$ in $A$ has no other intersections, as illustrated in Figure \ref{fi:case2-3}a. Removing this pseudoline from $A$ gives us the arrangement $A'$ which corresponds to $G'$. By induction, since $A'$ has one fewer crossing than $A$, $A'$ has a chord representation. This representation can be perturbed if necessary, so that the neighbor of $v$ is moved to the interior of the circle. Let $c$ be the chord of this representation on which $v$should lie. We can represent $A$ by placing $v$ on the crossing of $c$ with the circle and adding a new chord starting from $v$. By making the chord small enough, we can always do this without adding unwanted crossings. The resulting representation is shown in the rightmost image of Figure \ref{fi:case2-3}a.
\begin{figure}[b]
\centering
\includegraphics{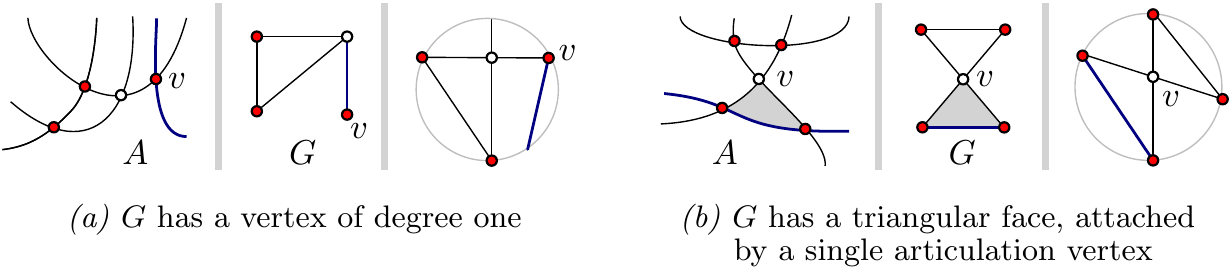}
\vspace{0.2em}
\caption{If $G$ has a vertex $v$ of degree one (a), perturb the the neighbor of $v$ to the interior of the circle and intersect the chord $v$ should be on with a new chord. If $G$ has a triangular face attached by a single articulation vertex $v$ (b), perturb $v$ to the interior of the circle and close the face by adding an extra chord.}
\label{fi:case2-3}
\end{figure}

\item If $G$ has a triangular face attached to the rest of the graph by a single articulation vertex $v$, removing the edge of this triangle opposite of $v$ gives us a graph $G'$ corresponding to an arrangement with two fewer crossings than $A$, which by induction can be represented as chords on a circle. We can perturb this representation if necessary, such that $v$ is in the interior of the circle. A representation of $A$ can now be drawn by adding two corner vertices where the chords crossing in $v$ intersect the circle, and connecting those with a new chord. This is illustrated in Figure \ref{fi:case2-3}b.

\item If $G$ has a bounded face $f$ with more than three sides, that shares at most one interior edge $e$ with another bounded face, form $G'$ by removing all edges (including vertices) of $f$ except $e$ from $G$. This corresponds to removing from $A$ all pseudolines that bound $f$ but no other bounded face to obtain $A'$. By induction, $A'$ has a chord representation, which can be perturbed so that the vertices of $f$ present in $A'$ are in the interior of the circle. The parts of $f$ that are already included in the representation of $A'$ consist at most of the shared edge, its endpoints, and two chords that should contain two more sides of $f$ (it is also possible for $A'$ to intersect $f$ in a single vertex, or not at all). By adding extra chords that cross these two chords near the points where they cross the circle, it is straightforward to extend the representation of $A'$ to a representation of $A$. This case is illustrated in Figure \ref{fi:case4}. If $f$ only shares one vertex with another closed face, the representation of $A'$ still includes two chords that should contain two sides of $f$ and the approach for completing the face is the same as before. If $A'$ does not intersect $f$ at all (this happens if $f$ does not share any vertex with another bounded face) we can draw the entire face as a chain of chords.
\begin{figure}[tb]
\centering
\includegraphics{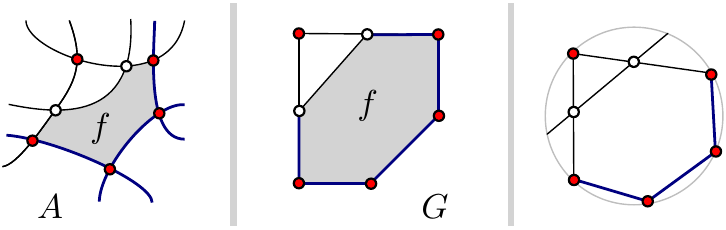}
\caption{If $G$ has a bounded face $f$ with more than three sides that shares at most one interior edge with another bounded face, perturb the shared vertices to the interior of the circle and close the face with a chain of chords.}
\label{fi:case4}
\end{figure}

\item\label{quads} In this case $G$ has an interior edge $e$, on one side of which are a quadrilateral $q$ and one or two triangles. One of the triangles shares the edge of $q$ that is not adjacent to $e$. The situation is illustrated in Figure \ref{fi:case5}. Form a smaller arrangement $A'$ by removing the pseudoline forming the side of $q$ opposite to $e$ (the bold, blue pseudoline in the figure). Note that this reduces the part of $G$ on that side of $e$ to a single triangle with one corner. Find a system of chords for $A'$ by induction. The removed pseudoline can now be inserted as a chord that intersects the circle close to the corner of the triangle, as is shown in the rightmost image of Figure \ref{fi:case5}. Nothing but the triangle can be crossed by this chord, since there is nothing else on this side of $e$ in the reduced arrangement $A'$. 
\begin{figure}[b]
\centering
\includegraphics{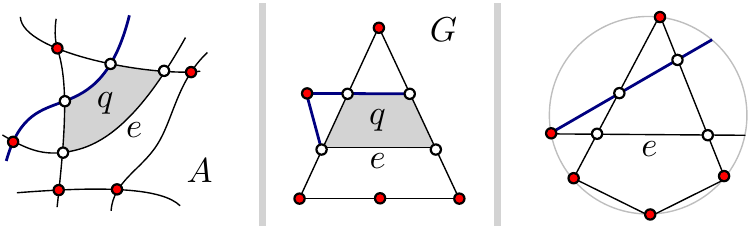}
\caption{If $G$ contains an interior edge $e$ adjacent to a quadrilateral $q$ that is in turn adjacent to one or two triangular faces, first represent $A'$ without the pseudoline forming the side of $q$ opposite of $e$, then insert the missing pseudoline as a chord.}
\label{fi:case5}
\end{figure}

\item Here we handle any remaining situation where none of the cases 1-5 apply. When we get to this case, we know that there must exist a face that is not a triangle, otherwise we would be in case 1. Let $r$ be any vertex of $G$ and let $f$ be a non-triangle face of $G$ that maximizes the distance (number of faces and bridge edges that must be crossed) between $r$ and $f$. This face $f$ has at most one edge $e$ that also has a non-triangle face on its other side, because if $f$ was incident to two or more non-triangle faces, one of them would be farther away from $r$ than $f$ itself. Therefore, all the other edges of $f$ can only have a triangle or nothing on their other side. If $f$ was connected to the rest of the graph through one of these triangles, we would be in case 3 (triangular face connected through a single articulation vertex), so it must be edge $e$ that has face $f$ and its attached triangles on one side and the rest of the graph on the other side. We also know that $f$ has at least five sides, otherwise we would be in case 5. Figure \ref{fi:case6} shows an example of this situation.     
\begin{figure}[tb]
\centering
\includegraphics{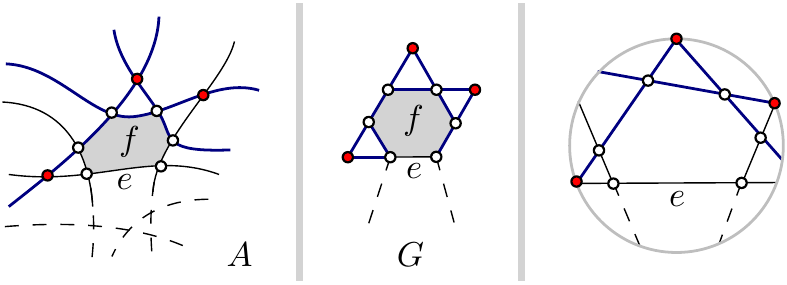}
\caption{If none of the other cases apply, $G$ contains a non-triangle face $f$ connected to the rest of the graph ($A'$) through edge $e$, with at most triangles attached to its other sides. First find a representation for $A'$, perturb the vertices of $e$ to the interior of the circle, and then insert chords for face $f$ and its incident triangles on the other side of $e$.}
\label{fi:case6}
\end{figure}

Let $G'$ be the graph formed from $G$ by removing all edges on the $f$-side of $e$.  bounding $f$, except the one that contain on $f$'s side of $e$ (except $e$ itself). By induction, the arrangement corresponding to $G'$ can be represented by chords. If the two ends of the pseudoline through $e$ bound triangles that were removed, place their apexes at the points where the corresponding chord crosses the circle, and place the apexes of the remaining triangles and endpoints of the remaining rays in $G$ in the appropriate order around the circle between these two points. Regardless of the precise placement of these points, the set of chords connecting them necessarily has the correct set of crossings to represent $G$. We can do this without introducing unwanted crossings, because in $G'$ there was nothing left on this side of $e$. The rightmost Figure \ref{fi:case6} shows the result of this step, the representation of the rest of the graph would be drawn below edge $e$. \hfill\qed
\end{enumerate}


\begin{cor}
Every weak outerplanar pseudoline arrangement can be represented by convex polygonal chains formed by two rays and one line segment, with at most two bends per chain.
\end{cor}

\noindent\textit{Proof.}\quad Find a representation by chords, perturb all corners to the interior of the circle, and add a ray at both ends of the chord, extending perpendicularly from the circle containing the chords to infinity. This extension cannot create any additional crossings.
\hfill\qed

\begin{cor}
Every weak outerplanar pseudoline arrangement can be represented by lines in the hyperbolic plane, or by semicircles with endpoints on a common line.
\end{cor}

\noindent\textit{Proof.}\quad These results follow by interpreting the circle and its chords as a Klein model of the hyperbolic plane, and then using the Poincar\'e model of the hyperbolic plane as a Euclidean halfplane and semicircles within it.
\hfill\qed

This result complements the fact that a weak arrangement with no $3$-clique can always be represented by hyperbolic lines, regardless of outerplanarity~\cite{BanCheEpp-SJDM-10}.

\section{Arbitrary arrangements}
\label{sec:arbitrary}
It is known that every (non-weak) pseudoline arrangement can be represented as a \emph{wiring diagram}~\cite{Goo-DM-80}, a system of monotone curves that lie on $n$ equally spaced horizontal lines except near a crossing, where the pseudolines follow straight line segments that connect one horizontal line to another. For weak pseudoline arrangements, such a representation is not fully general; for instance it cannot represent a set of three pseudolines that do not cross each other but all bound the same cell of the arrangement. However in this case a slightly more general representation is possible in which each pseudoline is a polygonal path with infinite downward vertical rays at both of its ends, and in which the endpoints of these rays are connected by a curve that stays on one of $n$ equally spaced horizontal lines except near a crossing as in a standard wiring diagram. We may assume that the vertical rays and crossings of this representation all have distinct $x$-coordinates.

Wiring diagrams represent arrangements by polygonal chains, but the chains are not generally convex. We show below how to transform these representations into a system of convex chains, while preserving the property that each chain has only a linear number of bends.

\begin{figure}[b]
  \centering
  \includegraphics{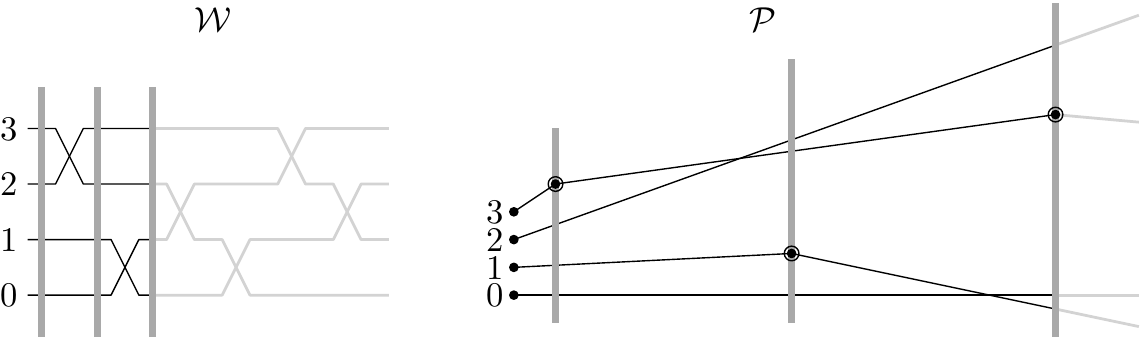}
  \caption{A wiring diagram $\W$ and part of the corresponding polyline representation $\P$}
  \label{fig:polyline-arrangement}
 \end{figure}

\begin{thm}
 Every $n$-element (non-weak) pseudoline arrangement can be drawn with convex polygonal chains, each of complexity at most $n$.
\end{thm}
\noindent\textit{Proof.}\quad
 We consider a wiring diagram $\W$ of the arrangement and construct a realization $\P$ with convex polygonal chains by sweeping $\W$ from left to right. For convenience we choose a wiring diagram $\W$ in which no two crossings occur one the same vertical line.
 We number the pseudolines from $0$ to $n-1$ according to their bottom-to-top order on the very left of $\W$. We define the polyline for $i$ to start at $(0,i)$ and emerge from there to the right with a slope of $s_i$, such that $0=s_0 < s_1 < \cdots < s_{n-1} = 1$. This is illustrated in Figure \ref{fig:polyline-arrangement}.

 At every intermediate time the rightmost segment of each polygonal chain is considered to be an infinite ray. To get two polygonal chains $i$ and $j$, say with $i>j$, crossing we shall later introduce a bend to the ray of the upper polygonal chain, which is $i$, giving the new ray a slightly smaller slope. The new slope is chosen to be smaller than the slope of the ray of $j$ but larger than all the current slopes of rays below $j$. Thus the rays of $i$ and $j$ cross and to the right of that crossing again all rays are divergent.

 More formally, we define the \emph{borderlines of the construction} to be a pair of a vertical line in $\W$ and a vertical line in the polygonal chain representation $\P$ such that
 \begin{itemize}
  \item between any two consecutive borderlines there is exactly one crossing,
  \item both representations to the left of the borderlines are equivalent
  \item and the polygonal chains to the right of the borderline in $\P$ are divergent rays.
 \end{itemize}
 The initial borderlines are placed immediately to the right of all left endpoints of pseudolines in both representations $\W$ and $\P$. A \emph{step} is a transition between the current borderline to the next borderline. In each step we consider the corresponding crossing in $\W$, say pseudoline $i$ crosses $j$ and we have $i > j$, i.e., on the current borderline pseudoline $i$ lies above pseudoline $j$ and on the next the order is swapped.

 In $\P$ we introduce a bend on the ray of polygonal chain $i$ at its intersection with the current borderline. This introduces a new finite segment and a new infinite ray. If $k$ denotes the polygonal chain immediately below $j$ on the current borderline, we set the slope of the new infinite ray of $i$ to the mean of the current slopes of $j$ and $k$. If no such $k$ exists, i.e., $j$ is currently the bottommost polygonal chain, we set the new slope for the ray of $i$ to any number strictly less than the current slope of $j$. Evidently, the new ray of $i$ and the current ray of $j$ will intersect somewhere on the right of the current borderline. We set the next borderline in $\P$ to the right of that crossing.

 It is immediate to check that the pair of next borderlines again satisfies the properties above and hence we can continue in this manner until having advanced past the rightmost borderline in $\W$. Chopping in $\P$ each infinite ray at its intersection with the corresponding rightmost borderline, we obtain an equivalent representation of the pseudoline arrangement with finite polygonal chains. Since along each polygonal chain the slopes of segments decrease, all polygonal chains are convex. Moreover, polygonal chain $i$ participates in exactly $i$ crossings with polygonal chains of smaller number. Thus polygonal chain $i$ has exactly $i$ bends, i.e., consists of exactly $i+1 \leq n$ segments.
\hfill\qed

For smooth curves composed of multiple circular arcs and straight line segments, Bekos et al.~\cite{BekKauKob-GD-12} defined the \emph{curve complexity} to be the maximum number of arcs and segments in a single curve. By replacing each bend of the above result by a small circular arc, one obtains a smooth convex representation of the arrangement with curve complexity $O(n)$.
As we now show, these bounds are optimal.

\begin{thm}
\label{thm:many-bends}
There exist simple arrangements of $n$ pseudolines that, when represented by polygonal chains require some pseudolines to have $\Omega(n)$ bends.
\end{thm}

\noindent\textit{Proof.}\quad
It is long known that there exists a simple arrangement of $9$ pseudolines that cannot be represented by straight lines~\cite{levi1926}.
We form an arrangement of $n$ pseudolines by determining the crossings of a system of monotone curves in left-to-right order. To the left of all crossings, we group the pseudolines into $n/9$ groups of $9$ pseudolines each. Within each group, we form a system of crossings so the group forms a non-stretchable arrangement. To the right of all of these crossings, we move the line now in the $i$th position of its group to its original starting position in the group $i$ steps above it (discarding the bottom $i-1$ groups, which no longer have $9$ lines assigned to them and also discarding the lines that in this way would move above the topmost group). This move causes each group to again contain $9$ lines that have not yet crossed.
We repeat the same process of forming non-stretchable arrangements within groups of $9$ lines and then moving lines to different groups, until all lines and all groups have been discarded.

\begin{figure}[t]%
\centering
\includegraphics[width=0.7\columnwidth]{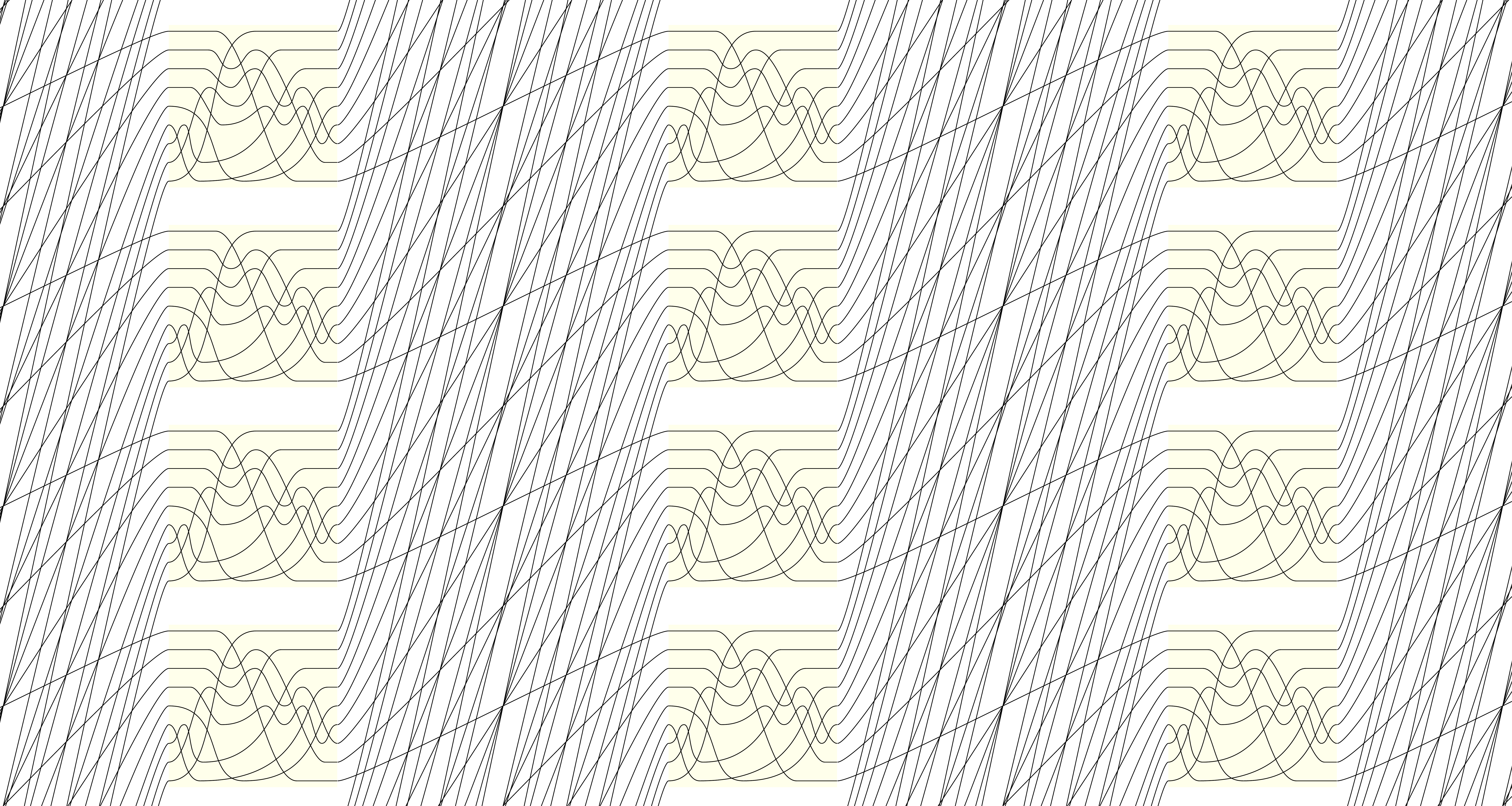}%
\caption{Part of an arrangement requiring quadratically many bends}%
\label{fi:many-bends}%
\vspace{-3mm}
\end{figure}

The result of this construction is a system of $\Omega(n^2)$ non-stretchable $9$-line arrangements (Figure~\ref{fi:many-bends}), each of which must contain at least one bend in its part of any realization of the whole arrangement. Thus the whole arrangement must have $\Omega(n^2)$ bends and there must be at least one pseudoline with $\Omega(n)$ bends.
\hfill\qed

\begin{thm}
There exist simple arrangements of $n$ pseudolines that, when represented by smooth piecewise-circular curves require some curves to have $\Omega(n)$ arcs.
\end{thm}

\noindent\textit{Proof.}\quad
Goodman and Pollack~\cite{GooPol-DCG-86} showed that the number of combinatorially distinct line arrangements is $2^{O(n\log n)}$; their proof uses only the facts that each line can be represented by a constant number of real-number coefficients and that the orientation of a triple of lines may be determined from the sign of a bounded number of polynomials in those coefficients. These facts are true also of circular arc arrangements, so the number of combinatorially distinct circular arc arrangements is again $2^{O(n\log n)}$ (with a different constant in the $O$-notation). In contrast, there are $2^{\Theta(n^2)}$ simple pseudoline arrangements~\cite{Fel-SoCG-96,Knu-AH-92}, a much larger number. Therefore, there exist  arrangements that cannot be realized by circular arcs. Choosing one of these arrangements and applying the same argument as Theorem~\ref{thm:many-bends} gives the result.
\hfill\qed

\section{Conclusions}

We have shown that every $n$-element simple pseudoline arrangement can be drawn with convex polygonal chains with at most $n$ bends, and that this is asymptotically best-possible. In contrast we presented an algorithm to draw every simple weak outerplanar arrangement with convex polygonal chains with at most $2$ bends. Since not every such arrangement is stretchable to a line arrangement this is almost best-possible. However, we also show that all such arrangements are stretchable in the hyperbolic plane.

\bibliographystyle{amsplain}
\bibliography{pseudolines}

\end{document}